\documentclass[prb,twocolumn,showpacs]{revtex4}
\usepackage{amsfonts}
\usepackage{amsmath}
\usepackage{dcolumn}
\usepackage{graphicx}
\usepackage{bm}

\begin{document}

\title{Class of exactly solvable $SO(n)$ symmetric spin chains with matrix
product ground states}
\author{Hong-Hao Tu and Guang-Ming Zhang}
\email{gmzhang@mail.tsinghua.edu.cn}
\affiliation{Department of Physics, Tsinghua University, Beijing 100084, China}
\author{Tao Xiang}
\affiliation{Institute of Physics, Chinese Academy of Sciences, Beijing 100190, China;\\
Institute of Theoretical Physics, Chinese Academy of Sciences, Beijing
100190, China}
\date{\today}

\begin{abstract}
We introduce a class of exactly solvable $SO(n)$ symmetric Hamiltonians with
matrix product ground states. For an odd $n\geq 3$ case, the ground state is
a translational invariant Haldane gap spin liquid state; while for an even $%
n\geq 4$ case, the ground state is a spontaneously dimerized state with
twofold degeneracy. In the matrix product ground states for both cases, we
identify a hidden antiferromagnetic order, which is characterized by
nonlocal string order parameters. The ground-state phase diagram of a
generalized $SO(n)$ symmetric bilinear-biquadratic model is discussed.
\end{abstract}

\pacs{75.10.Pq, 75.10.Jm, 03.65.Fd}
\maketitle

\section{Introduction}

One-dimensional quantum Heisenberg antiferromagnets have a long history and
show many fascinating properties. Since the Mermin-Wagner-Coleman theorem
\cite{Mermin-1966,Coleman-1973} forbids a continuous symmetry breaking in
one dimension, no classical N\'{e}el order can survive, even in zero
temperature. The rigorous solutions on particular models provide essential
insights to understand the properties of these quantum spin liquid states.
For instance, the spin-$1/2$ antiferromagnetic Heisenberg chain has a
Bethe-ansatz solution, \cite{Bethe-1931,Faddeev-1981} which yields a unique
spin singlet ground state, gapless spin-$1/2$ excitations and power-law
decay spin correlations. Meanwhile, additional next-nearest-neighbor
interactions can frustrate the nearest-neighbor antiferromagnetic
correlations. The Majumdar-Ghosh model \cite{Majumdar-1969} is such an
exactly solvable example, which has a twofold-degenerate dimerized ground
state, a finite-energy gap, and extremely short spin correlations.

Toward the quantum integer-spin models, Haldane gave a striking prediction
that an excitation gap occurs between the ground state and the excited
states. \cite{Haldane-1983} Although Haldane's argument is based on a
semiclassical large-$S$ expansion, it was later verified by numerical
studies for lower-$S$ cases. \cite%
{Nightingale-1986,Takahashi-1989,White-1993} Remarkably, Affleck, Kennedy,
Lieb, and Tasaki (AKLT) found a family of integer-spin chain Hamiltonians
with exact massive ground states, which are called valence bond solid (VBS)
states. \cite{Affleck-1987}\ The VBS states preserve spin rotational
symmetry, and exhibit exponentially decay spin correlations and gapped
excitations, thus, share the key features of Haldane gap spin liquid states
for the quantum integer-spin Heisenberg antiferromagnets. Although no true
long range order exists, den Nijs and Rommelse \cite{den Nijs-1989}\
observed a hidden antiferromagnetic order in the $S=1$ VBS state, and
introduced a set of nonlocal string order parameters to provide a faithful
quantification of the $S=1$ Haldane phase. The string order in $S=1$ VBS
state, Haldane gap, and the fourfold degeneracy in an open chain can be
understood by a hidden $Z_{2}\times Z_{2}$ symmetry breaking. \cite%
{Kennedy-1991,Oshikawa-1992,Suzuki-1995} However, a nonlocal string order
parameter\ that reflects correctly the hidden $Z_{S+1}\times Z_{S+1}$
symmetry of the higher-$S$ VBS states remains an open problem. \cite%
{Oshikawa-1992,Suzuki-1995,Schollwock-1996,Tu-AKLT}

Beside the studies on $SU(2)$-symmetric spin chains, quantum spin systems
with higher symmetry also attract much attention. For instance, the
Bethe-ansatz method for $SU(2)$ Heisenberg chains can be generalized to
models with $SU(n)$ symmetry. \cite{Sutherland-1975} It has been argued that
such an $SU(4)$-symmetric model can be achieved in electronic systems with
two-fold orbital degeneracy at quarter filling. \cite{YQLi-1998} Meanwhile,
Affleck \textit{et al}. \cite{Affleck-1991}\ first discussed the extension
of VBS states to $SU(2n)$-invariant extended VBS states, which break lattice
translational symmetry and charge-conjugation symmetry but remain invariant
under the combined operation of these two symmetries. Furthermore, Greiter
\textit{et al.} \cite{Greiter-2007}\ studied the $SU(n)$ spin chains with
exact valence bond solid ground states. Along with the rise of cold atomic
physics in optical lattices, Chen \textit{et al.} \cite{YPWang-2005}\
constructed an $SU(4)$ Majumdar-Ghosh model with exact plaquette ground
states by using spin-$3/2$ fermions. Very recently, Arovas \cite{Arovas-2008}%
\ explored a family of novel $SU(n)$ simplex solid states, which are natural
generalizations of $SU(2)$ VBS states of AKLT models. Besides these models
with $SU(n)$ symmetry, Schuricht and Rachel \cite{Rachel-2008}\ considered $%
Sp(2n)$ VBS states and their parent Hamiltonians.

In this paper, we will introduce a class of $SO(n)$-symmetric Hamiltonians
with matrix product states as their exact ground states. However, these $%
SO(n)$ symmetric spin chains show a different even-odd effect. For an odd $%
n=2l+1$, a periodic chain has a unique ground state. All these $SO(2l+1)$
matrix product states have a hidden antiferromagnetic order, which is
characterized by string order parameters. The nonlocal unitary
transformations are designed to explicitly reveal a hidden $(Z_{2}\times
Z_{2})^{l}$ symmetry. The breaking of this symmetry is responsible for the
Haldane gap, nonvanishing string order parameters, and $4^{l}$-fold
degeneracy in an open chain. However, for an even $n=2l$, a periodic chain
has a twofold dimerized ground state, which breaks translational symmetry.
Nevertheless, these $SO(2l)$ matrix product states also\ contain a hidden
antiferromagnetic order. Finally, the ground-state phase diagram of a
generalized $SO(n)$ symmetric bilinear-biquadratic model is obtained.

This paper is organized as follows. In Sec. II, the $SO(n)$ algebra and the
exactly solvable $SO(n)$ symmetric models will be introduced. In Sec. III,
the exact matrix product ground state of the $SO(n)$ model with $n=2l+1$
will be studied, in particular, with the examples of $n=3$ and $5$. The
hidden order in all these $SO(2l+1)$ matrix product states and the
corresponding hidden symmetry are identified. Section IV is devoted to an
analysis of the $SO(n)$ model with $n=2l$, which has a dimerized ground
state breaking lattice translational symmetry. In Sec. V, a generalized $%
SO(n)$ symmetric bilinear-biquadratic model is introduced and their
ground-state properties are discussed in detail. A conclusion is presented
in Sec. VI.

\section{Model Hamiltonian}

Let us begin with a one-dimensional $SO(n)$ chain of $N$ lattice sites ($N$
even). On each site, the local Hilbert space $\mathcal{%
%TCIMACRO{\U{2102} }%
%BeginExpansion
\mathbb{C}
%EndExpansion
}^{n}$ contains $n$ states $\left\vert n^{a}\right\rangle \ (1\leq a\leq n)$%
, which can be rotated within the $SO(n)$ space via the following vector
relations:
\begin{equation}
L^{ab}|n^{c}\rangle =i\delta _{bc}|n^{a}\rangle -i\delta _{ac}|n^{b}\rangle ,
\label{Rotation}
\end{equation}%
where $L^{ab}$ $(a<b)$ are the $n(n-1)/2$ generators of the $SO(n)$ Lie
algebra. The vector relations constitute the $n$-dimensional representation
of $SO(n)$ algebra and the following commutation relations\ hold: \cite%
{Georgi-1999}
\begin{equation}
\lbrack L^{ab},L^{cd}]=i\left( \delta _{ad}L^{bc}+\delta _{bc}L^{ad}-\delta
_{ac}L^{bd}-\delta _{bd}L^{ac}\right) .
\end{equation}

According to the Lie algebra, the tensor product of two $SO(n)$ vectors can
be decomposed as a direct sum of an $SO(n)$ singlet with a dimension $1$, an
antisymmetric $SO(n)$ tensor with a dimension $n(n-1)/2$, and a symmetric $%
SO(n)$ tensor with a dimension $(n+2)(n-1)/2$, i.e.,
\begin{equation}
\underline{n}\otimes \underline{n}=\underline{1}\oplus \underline{n(n-1)/2}%
\oplus \underline{(n+2)(n-1)/2},  \label{eq:decom}
\end{equation}%
where the number above each underline is the dimension of the corresponding
irreducible representation. For $n=3$, we recover the well-known
Clebsch-Gordan decomposition $\underline{3}\otimes \underline{3}=\underline{1%
}\oplus \underline{3}\oplus \underline{5}$ of two spin-$1$ representations.
According to the $SO(n)$ decomposition scheme (\ref{eq:decom}), the wave
functions in each irreducible representation channel can be obtained
explicitly. The maximally entangled $SO(n)$ singlet wave function is written
as $\frac{1}{\sqrt{n}}\sum_{a=1}^{n}\left\vert n^{a}\right\rangle
_{i}\left\vert n^{a}\right\rangle _{j}$, and the wave functions of the
antisymmetric channel are expressed as
\begin{equation}
\frac{1}{\sqrt{2}}\left( \left\vert n^{a}\right\rangle _{i}\left\vert
n^{b}\right\rangle _{j}-\left\vert n^{b}\right\rangle _{i}\left\vert
n^{a}\right\rangle _{j}\right) .
\end{equation}%
Finally, the symmetric channel contains $n(n-1)/2$ states with the wave
functions
\begin{equation}
\frac{1}{\sqrt{2}}\left( \left\vert n^{a}\right\rangle _{i}\left\vert
n^{b}\right\rangle _{j}+\left\vert n^{b}\right\rangle _{i}\left\vert
n^{a}\right\rangle _{j}\right) ,
\end{equation}%
and the rest of $n-1$ states with the wave functions
\begin{equation}
\frac{1}{\sqrt{2}}\left( \left\vert n^{a}\right\rangle _{i}\left\vert
n^{a}\right\rangle _{j}-\left\vert n^{b}\right\rangle _{i}\left\vert
n^{b}\right\rangle _{j}\right) .
\end{equation}

For the three $SO(n)$ channels given in Eq. (\ref{eq:decom}), the bond
Casimir charge $\sum_{a<b}(L_{i}^{ab}+L_{j}^{ab})^{2}$ for two adjacent
sites takes the values $0$, $2n-4$, and $2n$, respectively. Together with
the single-site Casimir charge $\sum_{a<b}(L_{i}^{ab})^{2}=n-1$, one can
write the $SO(n)$ symmetric bilinear interaction term as a polynomial of
bond projection operators,%
\begin{eqnarray}
\sum_{a<b}L_{i}^{ab}L_{j}^{ab} &=&(1-n)\mathcal{P}_{\underline{1}}(i,j)-%
\mathcal{P}_{\underline{n(n-1)/2}}(i,j)  \notag \\
&&+\mathcal{P}_{\underline{(n+2)(n-1)/2}}(i,j),  \label{eq:projector}
\end{eqnarray}%
where the bond projectors $\mathcal{P}_{\underline{1}}(i,j)$, $\mathcal{P}_{%
\underline{n(n-1)/2}}(i,j)$, and $\mathcal{P}_{\underline{(n+2)(n-1)/2}%
}(i,j) $ project the states of two adjacent sites $i$ and $j$ onto the three
$SO(n)$ channels in Eq. (\ref{eq:decom}), respectively. Using the property
of projection operators, we square Eq. (\ref{eq:projector}) and obtain the $%
SO(n)$ symmetric biquadratic interaction term as
\begin{eqnarray}
\left( \sum_{a<b}L_{i}^{ab}L_{j}^{ab}\right) ^{2} &=&(n-1)^{2}\mathcal{P}_{%
\underline{1}}(i,j)+\mathcal{P}_{\underline{n(n-1)/2}}(i,j)  \notag \\
&&+\mathcal{P}_{\underline{(n+2)(n-1)/2}}(i,j).
\end{eqnarray}%
Combined with the completeness relation of the projectors,%
\begin{equation}
\mathcal{P}_{\underline{1}}(i,j)+\mathcal{P}_{\underline{n(n-1)/2}}(i,j)+%
\mathcal{P}_{\underline{(n+2)(n-1)/2}}(i,j)=1,
\end{equation}%
we can express the bond projection operators with the $SO(n)$ generators as%
\begin{eqnarray}
\begin{pmatrix}
\mathcal{P}_{\underline{1}}(i,j) \\
\mathcal{P}_{\underline{n(n-1)/2}}(i,j) \\
\mathcal{P}_{\underline{(n+2)(n-1)/2}}(i,j)%
\end{pmatrix}
&=&%
\begin{pmatrix}
\frac{-1}{n(n-2)} & 0 & \frac{1}{n(n-2)} \\
\frac{n-1}{2(n-2)} & \frac{-1}{2} & \frac{-1}{2(n-2)} \\
\frac{n-1}{2n} & \frac{1}{2} & \frac{1}{2n}%
\end{pmatrix}
\notag \\
&&\times
\begin{pmatrix}
1 \\
\sum_{a<b}L_{i}^{ab}L_{j}^{ab} \\
(\sum_{a<b}L_{i}^{ab}L_{j}^{ab})^{2}%
\end{pmatrix}%
.  \label{eq:so(n)projector}
\end{eqnarray}

Now we define our model Hamiltonian as
\begin{equation}
H_{SO(n)}=\sum_{i}\mathcal{P}_{\underline{(n+2)(n-1)/2}}(i,i+1),
\label{eq:soham}
\end{equation}%
which is a bilinear-biquadratic Hamiltonian in terms of the $SO(n)$
generators according to Eq. (\ref{eq:so(n)projector}). This model has exact
matrix product ground states, which will be extensively studied below.
Although the exact excited states are not known, we argue that there is a
finite energy gap above the ground states. For a projector Hamiltonian such
as Eq. (\ref{eq:soham}), this argument can be proved rigorously using a
method proposed by Knabe, \cite{Knabe-1988}\ who found that the lower bounds
of energy gaps of infinite systems can be obtained by diagonalizing
finite-size systems.

In Secs. III and IV, we will discuss odd $n=2l+1$ and even $n=2l$ cases
separately because the nature of the matrix product ground state depends on
the parity of $n$. Mathematically speaking, the $SO(2l+1)$ and $SO(2l)$
algebras are quite different. According to the Cartan classification scheme,
\cite{Georgi-1999}\ the $SO(2l+1)$ algebra belongs to $B_{l}$ type, while
the $SO(2l)$ algebra are $D_{l}$ type.

\section{Odd-$n$ case}

Let us assume $n=2l+1$, where $l$ is an integer ($l\geq 1$). To achieve the
exact ground state of model Hamiltonian (\ref{eq:soham}), one has to resort
to a fascinating property of $SO(n)$ algebra --- the spinor representation.
An elegant way to construct the spinor representation of $SO(2l+1)$ algebra
is to introduce $(2l+1)$ gamma matrices satisfying the Clifford algebra $%
\{\Gamma ^{a},\Gamma ^{b}\}=2\delta _{ab}$. Then an \textit{irreducible}
spinor representation of $SO(2l+1)$ is immediately constructed by $\Gamma
^{ab}=[\Gamma ^{a},\Gamma ^{b}]/2i$. The product of $\Gamma ^{a}$ and $%
\Gamma ^{b}$ can be expressed as $\Gamma ^{a}\Gamma ^{b}=\delta
_{ab}+i\Gamma ^{ab}$. For each lattice site $j$, if the following matrix
state is introduced:
\begin{equation}
g_{j}=\sum_{a=1}^{2l+1}\Gamma ^{a}\left\vert n^{a}\right\rangle _{j},
\label{eq:gmatrix1}
\end{equation}%
then the bond product of $g_{j}$ at any two neighboring sites $g_{j}g_{j+1}$
is given by%
\begin{eqnarray}
g_{j}g_{j+1} &=&i\sum_{a<b}\Gamma ^{ab}\left( \left\vert n^{a}\right\rangle
_{j}\left\vert n^{b}\right\rangle _{j+1}-\left\vert n^{b}\right\rangle
_{j}\left\vert n^{a}\right\rangle _{j+1}\right)  \notag \\
&&+\sum_{a}\left\vert n^{a}\right\rangle _{j}\left\vert n^{a}\right\rangle
_{j+1},
\end{eqnarray}%
where the first two terms belongs to the antisymmetric channel in Eq. (\ref%
{eq:decom}) and the latter term is the $SO(2l+1)$ singlet. Since $\mathcal{P}%
_{\underline{(n+2)(n-1)/2}}(j,j+1)$ performs the projection onto the states
of the symmetric $\underline{(n+2)(n-1)/2}$ channel, the matrix product
state defined by
\begin{eqnarray}
|\Psi \rangle &=&\mathrm{Tr}(g_{1}g_{2}\ldots g_{N})  \notag \\
&=&\sum_{a_{1}...a_{N}}\mathrm{Tr}(\Gamma ^{a_{1}}\Gamma ^{a_{2}}\ldots
\Gamma ^{a_{N}})\left\vert n^{a_{1}}n^{a_{2}}\ldots n^{a_{N}}\right\rangle ,
\label{eq:MPS}
\end{eqnarray}%
is always the zero energy ground state of the Hamiltonian (\ref{eq:soham})
in a periodic boundary condition. This state preserves $SO(2l+1)$ symmetry
and lattice translational symmetry. For an open chain, there are totally $%
4^{l}$-degenerate ground states, which can be distinguished by their edge
states.

To compute the correlation functions in the matrix product ground state, we
set up a transfer matrix method \cite{Klumper-1991}\ by introducing%
\begin{equation}
G_{A}=\sum_{ab}\left\langle n^{a}|A|n^{b}\right\rangle \left( \bar{\Gamma}%
^{a}\otimes \Gamma ^{b}\right) ,
\end{equation}%
where $A$ is an operator acting on a single site and $\bar{\Gamma}^{a}$
denotes the complex conjugate of $\Gamma ^{a}$. Specifically, the transfer
matrix $G$ is written as $G=\sum_{a}\bar{\Gamma}^{a}\otimes \Gamma ^{a}$.
Then a two-point correlation function in thermodynamic limit can be written
as%
\begin{equation}
\left\langle L_{i}^{ab}L_{j}^{ab}\right\rangle =\lim_{N\rightarrow \infty }%
\frac{\mathrm{Tr}[(G)^{N-j+i-1}G_{ab}(G)^{j-i-1}G_{ab}]}{\mathrm{Tr}(G)^{N}},
\end{equation}%
where $G_{ab}=i(\bar{\Gamma}^{a}\otimes \Gamma ^{b}-\bar{\Gamma}^{b}\otimes
\Gamma ^{a})$. In a long distant limit, the two-point correlation functions
of $SO(2l+1)$ generators decay exponentially as%
\begin{equation}
\left\langle L_{i}^{ab}L_{j}^{ab}\right\rangle \sim \exp \left( -\frac{|j-i|%
}{\xi }\right) ,  \label{eq:decay}
\end{equation}%
with the correlation length $\xi =1/\ln \left\vert \frac{2l+1}{2l-3}%
\right\vert $.

Here we note that the $SO(2l+1)$ symmetric model has a deep relation with
the quantum integer-spin chains. On each lattice site, the $(2l+1)$ vectors
of $SO(2l+1)$ can be constructed from the $S=l$ quantum spin states. In the $%
SU(2)$ spin language, the last two channels in Eq. (\ref{eq:decom}) for odd $%
n=2l+1$ correspond to the total bond spin $S=1,3,\ldots ,2l-1$ and $%
S=2,4,\ldots ,2l$ states, respectively. In other words, the $SO(2l+1)$ bond
projection operators can be expressed using the spin projection operators $%
P_{S=m}$ as
\begin{eqnarray}
\mathcal{P}_{\underline{2l^{2}+l}}(i,j) &=&\sum_{m=1}^{l}P_{S=2m-1}(i,j), \\
\mathcal{P}_{\underline{2l^{2}+3l}}(i,j) &=&\sum_{m=1}^{l}P_{S=2m}(i,j).
\end{eqnarray}%
Thus, the role of $\mathcal{P}_{\underline{2l^{2}+3l}}(i,j)$ is to project
onto nonzero even total spin states. Based on this property, we can further
show that the matrix product wave function (\ref{eq:MPS}) is also the ground
state of the following quantum integer-spin Hamiltonian:
\begin{equation}
H_{SU(2)}=\sum_{i}\sum_{m=1}^{l}J_{m}P_{S=2m}(i,i+1),  \label{eq:model2}
\end{equation}%
with all $J_{m}>0$. This model can be written as a polynomial of nearest
neighbor spin exchange interactions $\mathbf{S}_{i}\cdot \mathbf{S}_{i+1}$
up to $2l$ powers and is therefore $SU(2)$-invariant. However, the ground
state (\ref{eq:MPS}) possesses an \textit{emergent} $SO(2l+1)$ symmetry.

It is interesting to compare $H_{SU(2)}$ with the AKLT model of valence bond
solid proposed by Affleck \textit{et al}., \cite{Affleck-1987,Arovas-1988}
\begin{equation}
H_{\text{\textrm{AKLT}}}=\sum_{i}\sum_{m=l+1}^{2l}K_{m}P_{S=m}(i,i+1),
\end{equation}%
with all $K_{m}>0$. The ground state of $H_{\mathrm{AKLT}}$ is also a matrix
product state similar to Eq. (\ref{eq:MPS}), but the local $g$ matrix for
AKLT model is now a $(S+1)\times (S+1)=(l+1)\times (l+1)$ matrix. \cite%
{Suzuki-1995} When $l=1$, both models $H_{SO(n)}$ and $H_{SU(2)}$ become
exactly the same as the $S=1$ AKLT model $H_{\mathrm{AKLT}}$, whose ground
state is the celebrated $S=1$ VBS state. When $l>1$, we emphasis that $%
H_{SU(2)}$ and $H_{\mathrm{AKLT}}$ differ from each other. In Secs. III B
and III C, we will show that their matrix product ground states have very
different hidden structures and belong to different topological phases,
although they both belong to the Haldane liquid states.

\subsection{$SO(3)$ matrix product state: $S=1$ VBS}

In order to investigate the property of the $SO(2l+1)$ matrix product state,
we briefly review the $SO(3)$-symmetric $S=1$ VBS state as a warm up. In
this case, the $SO(3)$ vectors can be represented by the $S=1$ spin states,
\begin{equation}
\left\vert n^{1}\right\rangle =\frac{1}{\sqrt{2}}(\left\vert -1\right\rangle
-\left\vert 1\right\rangle ),\left\vert n^{2}\right\rangle =\frac{i}{\sqrt{2}%
}(\left\vert 1\right\rangle +\left\vert -1\right\rangle ),\left\vert
n^{3}\right\rangle =\left\vert 0\right\rangle .
\end{equation}%
and the $SO(3)$ generators are defined by spin-$1$ operators as $%
L^{12}=-S^{z},$ $L^{13}=S^{y},$ and $L^{23}=-S^{x}$. Moreover, the Clifford
algebra is satisfied by the Pauli matrices as $\{\sigma ^{a},\sigma
^{b}\}=2\delta _{ab}$. According to Eq. (\ref{eq:gmatrix1}), the local $g$
matrix can be written as%
\begin{equation}
g_{j}=%
\begin{pmatrix}
\left\vert 0\right\rangle _{j} & \sqrt{2}\left\vert -1\right\rangle _{j} \\
-\sqrt{2}\left\vert 1\right\rangle _{j} & -\left\vert 0\right\rangle _{j}%
\end{pmatrix}%
,
\end{equation}%
which generates the matrix form of the $S=1$ VBS state. Although the
two-point spin-correlation functions in this state decay exponentially as
shown in Eq. (\ref{eq:decay}), it has been observed that the upper and down
spins lie alternately along the lattice, sandwiched by arbitrary number of
non-polarized spin states. This hidden diluted antiferromagnetic order can
be characterized by a nonlocal string order parameter first proposed by den
Nijs and Rommelse, \cite{den Nijs-1989}
\begin{equation}
\mathcal{O}^{\mu }=\lim_{\left\vert j-i\right\vert \rightarrow \infty
}\langle S_{i}^{\mu }\prod_{r=i}^{j-1}\exp (i\pi S_{r}^{\mu })S_{j}^{\mu
}\rangle =\frac{4}{9},\,
\end{equation}%
where $\mu =x$, $y$, or $z$.

\begin{figure}[tbp]
\includegraphics[scale=0.5]{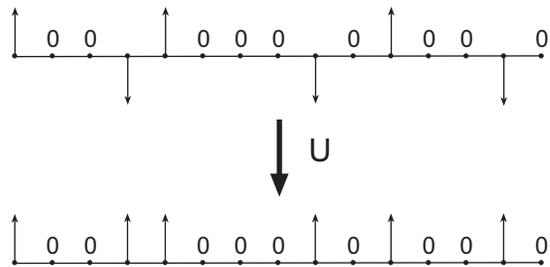}
\caption{Kennedy-Tasaki unitary transformation defined in Eq. (\protect\ref%
{eq:KennedyTasaki}) for a typical configuration of the spin-$1$ VBS states.
The hidden antiferromagnetic order is transformed to a dilute ferromagnetic
order.}
\label{fg:KennedyTasaki}
\end{figure}

On the other hand, the $S=1$ VBS states on a finite open chain have two
nearly free $S=1/2$ edge degrees of freedom at the end of the chain and are
thus fourfold degenerate. Both the hidden string order and the degeneracy in
an open chain can be understood as natural consequences of a hidden $%
Z_{2}\times Z_{2}$ symmetry breaking. To manifest the hidden symmetry, the
key is a nonlocal unitary transformation defined by \cite%
{Kennedy-1991,Oshikawa-1992}
\begin{equation}
U=\prod_{j<i}\exp (i\pi S_{j}^{z}S_{i}^{x}).  \label{eq:KennedyTasaki}
\end{equation}%
In the standard $S^{z}$ representation, $\exp \left( i\pi S^{x}\right) $
flips $\left\vert m\right\rangle $ to $\left\vert -m\right\rangle $ $(m=\pm
1,0)$ and multiplies the state with a phase factor $(-1)$. The physical
meaning of the unitary transformation $U$ can be explained as follows. For a
given spin configuration on a finite open chain, all $\left\vert
0\right\rangle $ are left alone and we look for the non-zero spins from the
left to the right. Suppose there is a non-zero spin at site $i$, we count
the number of $\left\vert 1\right\rangle $ and $\left\vert -1\right\rangle $
on the sites to the left of site $i$. If the number is even, we left the
spin at site $i$ unchanged. If the number is odd, we flip the $i$-site spin.
Finally, an additional phase factor $(-1)$ may be taken into account,
depending on the total site number and each spin configuration. An example
of the unitary transformation $U$ on a typical configuration of the spin-$1$
VBS state is shown in Fig. \ref{fg:KennedyTasaki}.

When applying the Kennedy-Tasaki unitary transformation to the Hamiltonian,
the $SO(3)$-symmetric AKLT model is transformed to a model with a discrete $%
Z_{2}\times Z_{2}$ symmetry, \cite{Kennedy-1991,Oshikawa-1992} and two of
the den Nijs-Rommelse string order parameters become the usual two-point
spin-correlation functions. Thus, the nonvanishing string order parameter
measures the hidden $Z_{2}\times Z_{2}$ symmetry breaking of the original
model. The breaking of this hidden discrete symmetry leads to the opening of
the Haldane gap, the hidden antiferromagnetic order, and the fourfold
degeneracy in an open chain and thus provides a unified explanation of the
exotic features in $S=1$ VBS states.

\subsection{$SO(5)$ matrix product state: A projected valence bond solid}

The next example is the $SO(5)$-symmetric matrix product state with $l=2$.
Actually, Scalapino \textit{et al}. \cite{Scalapino-1998}\ proposed this
state to describe the $SO(5)$ \textquotedblleft superspin\textquotedblright\
phase on a ladder system of interacting electrons. Here, it is convenient to
introduce the $SO(5)$ vectors by means of the $S=2$ states,
\begin{eqnarray}
|n^{1}\rangle &=&\frac{i}{\sqrt{2}}(\left\vert -2\right\rangle -\left\vert
2\right\rangle ),|n^{2}\rangle =\frac{1}{\sqrt{2}}(\left\vert 2\right\rangle
+\left\vert -2\right\rangle ),  \notag \\
|n^{3}\rangle &=&\frac{1}{\sqrt{2}}(\left\vert -1\right\rangle -\left\vert
1\right\rangle ),|n^{4}\rangle =\frac{i}{\sqrt{2}}(\left\vert 1\right\rangle
+\left\vert -1\right\rangle ),  \notag \\
|n^{5}\rangle &=&\left\vert 0\right\rangle .
\end{eqnarray}%
Moreover, we define the $SO(5)$ gamma matrices as
\begin{eqnarray}
\Gamma ^{1} &=&\sigma ^{2}\otimes \sigma ^{0},\text{ }\Gamma ^{2}=\sigma
^{1}\otimes \sigma ^{0},\text{ }\Gamma ^{3}=\sigma ^{3}\otimes \sigma ^{1},
\notag \\
\Gamma ^{4} &=&\sigma ^{3}\otimes \sigma ^{2},\text{ }\Gamma ^{5}=\sigma
^{3}\otimes \sigma ^{3}.
\end{eqnarray}%
Then the local $g$ matrix can be written as%
\begin{equation}
g_{j}=%
\begin{pmatrix}
\left\vert 0\right\rangle _{j} & \sqrt{2}\left\vert -1\right\rangle _{j} &
\sqrt{2}\left\vert -2\right\rangle _{j} & 0 \\
-\sqrt{2}\left\vert 1\right\rangle _{j} & -\left\vert 0\right\rangle _{j} & 0
& \sqrt{2}\left\vert -2\right\rangle _{j} \\
\sqrt{2}\left\vert 2\right\rangle _{j} & 0 & -\left\vert 0\right\rangle _{j}
& -\sqrt{2}\left\vert -1\right\rangle _{j} \\
0 & \sqrt{2}\left\vert 2\right\rangle _{j} & \sqrt{2}\left\vert
1\right\rangle _{j} & \left\vert 0\right\rangle _{j}%
\end{pmatrix}%
.  \label{eq:gmatrix2}
\end{equation}

\begin{figure}[tbp]
\includegraphics[scale=0.8]{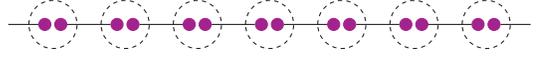}
\caption{(Color online) The schematic of a projected $SO(5)$ VBS state. Each
dot denotes a spin-$3/2$ fermion. The solid lines represent $SO(5)$ singlet
valence bond, and the dashed circles indicate the projection of two spin-$%
3/2 $ fermions to form spin-$2$ site-quintet states.}
\label{VBSFig}
\end{figure}

In fact, the $SO(5)$ matrix product state can be interpreted as a projected $%
SO(5)$\ VBS state (Fig. \ref{VBSFig}). By using two spin-$3/2$ fermions, the
spin-$2$ states can be constructed as \cite{Wu-2003,Tu-2006}
\begin{eqnarray}
\left\vert 2\right\rangle &=&\psi _{\frac{3}{2}}^{\dag }\psi _{\frac{1}{2}%
}^{\dag }\left\vert \mathrm{vac}\right\rangle ,\text{ \ }\left\vert
-2\right\rangle =\psi _{-\frac{1}{2}}^{\dag }\psi _{-\frac{3}{2}}^{\dag
}\left\vert \mathrm{vac}\right\rangle ,  \notag \\
\left\vert 1\right\rangle &=&\psi _{\frac{3}{2}}^{\dag }\psi _{-\frac{1}{2}%
}^{\dag }\left\vert \mathrm{vac}\right\rangle ,\text{ }\left\vert
-1\right\rangle =\psi _{\frac{1}{2}}^{\dag }\psi _{-\frac{3}{2}}^{\dag
}\left\vert \mathrm{vac}\right\rangle ,  \notag \\
\left\vert 0\right\rangle &=&\frac{1}{\sqrt{2}}(\psi _{\frac{3}{2}}^{\dag
}\psi _{-\frac{3}{2}}^{\dag }+\psi _{\frac{1}{2}}^{\dag }\psi _{-\frac{1}{2}%
}^{\dag })\left\vert \mathrm{vac}\right\rangle ,
\end{eqnarray}%
where $\psi _{\alpha }^{\dag }$ creates a fermion with spin components $%
\alpha =\pm 3/2,\pm 1/2$. Because only site quintet ($S=2$) and site singlet
($S=0$) are allowed for two spin-$3/2$ fermions on a single site, an extra
projection has to be implemented to remove the site-singlet state. Owing to $%
SO(5)\simeq Sp(4)$, there exists an antisymmetric matrix $\mathcal{R}%
=i\sigma ^{1}\otimes \sigma ^{2}$ with the following properties:%
\begin{eqnarray}
\mathcal{R}^{2} &=&-1,\text{ }\mathcal{R}^{\dagger }=\mathcal{R}^{-1}=%
\mathcal{R}^{T}=-\mathcal{R},  \notag \\
\mathcal{R}\Gamma ^{a}\mathcal{R}^{-1} &=&(\Gamma ^{a})^{T},\text{ }\mathcal{%
R}\Gamma ^{ab}\mathcal{R}^{-1}=-(\Gamma ^{ab})^{T}.
\end{eqnarray}%
Using the $\mathcal{R}$ matrix, the $SO(5)$ matrix product state in a
periodic chain can be written in a projected VBS wave function as%
\begin{equation}
\left\vert \Psi _{SO(5)}\right\rangle =\prod_{j}\mathcal{P}%
_{S=2}(j)(\sum_{\alpha \beta }\psi _{j,\alpha }^{\dag }\mathcal{R}_{\alpha
\beta }\psi _{j+1,\beta }^{\dag })\left\vert \mathrm{vac}\right\rangle ,
\end{equation}%
where $\mathcal{P}_{S=2}(j)$ is the site-quintet projector and $\sum_{\alpha
\beta }\psi _{j,\alpha }^{\dag }\mathcal{R}_{\alpha \beta }\psi _{j+1,\beta
}^{\dag }$ is an $SO(5)$-invariant valence bond singlet. For an\ open
boundary condition, the chain is ended with two nearly free spin-$3/2$
degrees of freedom leading to $16$ degenerate ground states. Here we recall
that the edge states of the $S=2$ VBS states of the AKLT model are spin-$1$
degrees of freedom, which are sharply different from our $SO(5)$ matrix
product states.

Similar to the spin-$1$ VBS state, the $SO(5)$ matrix product states have an
interesting hidden string order. Since the $SO(5)$ algebra is rank $2$, one
can classify the states by using two quantum numbers (weights) corresponding
to the mutual commuting Cartan generators $L^{12}$ and $L^{34}$ as
\begin{eqnarray}
L^{12}\left\vert m_{1},m_{2}\right\rangle &=&m_{1}\left\vert
m_{1},m_{2}\right\rangle ,  \notag \\
L^{34}\left\vert m_{1},m_{2}\right\rangle &=&m_{2}\left\vert
m_{1},m_{2}\right\rangle .
\end{eqnarray}%
These states, characterized by the $SO(5)$ weights, are related to those
states denoted by the usual $S^{z}$ quantum numbers as follows:
\begin{eqnarray}
\left\vert 1,0\right\rangle &=&\left\vert 2\right\rangle ,\text{ }\left\vert
-1,0\right\rangle =\left\vert -2\right\rangle ,\text{ }\left\vert
0,0\right\rangle =\left\vert 0\right\rangle ,  \notag \\
\left\vert 0,-1\right\rangle &=&\left\vert 1\right\rangle ,\text{ \ \ }%
\left\vert 0,1\right\rangle =\left\vert -1\right\rangle .
\end{eqnarray}%
When we define
\begin{equation}
\Gamma ^{\pm }=\frac{1}{2}\left( \Gamma ^{3}\pm i\Gamma ^{4}\right) ,\text{ }%
\Gamma _{\pm }=\frac{1}{2}\left( \Gamma ^{2}\pm i\Gamma ^{1}\right) ,
\end{equation}%
the local $g$ matrix in Eq. (\ref{eq:gmatrix2}) can be rewritten as%
\begin{eqnarray}
g_{j} &=&\sqrt{2}\Gamma ^{+}\left\vert 0,1\right\rangle _{j}-\sqrt{2}\Gamma
^{-}\left\vert 0,-1\right\rangle _{j}+\sqrt{2}\Gamma _{+}\left\vert
-1,0\right\rangle _{j}  \notag \\
&&+\sqrt{2}\Gamma _{-}\left\vert 1,0\right\rangle _{j}+\Gamma ^{5}\left\vert
0,0\right\rangle _{j}.
\end{eqnarray}%
By considering the property of Clifford algebra, it can be found that $%
\left\vert 1,0\right\rangle $ and $\left\vert -1,0\right\rangle $ must
appear alternately in the $SO(5)$ matrix product states despite arbitrary
numbers of $\left\vert 0,0\right\rangle $ and $\left\vert 0,\pm
1\right\rangle $ between them. At the same time, $\left\vert
0,1\right\rangle $ and $\left\vert 0,-1\right\rangle $ also appear
alternately with arbitrary numbers of $\left\vert 0,0\right\rangle $ and $%
\left\vert \pm 1,0\right\rangle $ between them. For example, a typical
configuration of the $SO(5)$ matrix product state is
\begin{equation*}
\begin{array}{crcccccccccccccccl}
m_{1}: & \quad \cdots & 0 & \uparrow & 0 & 0 & \downarrow & \uparrow & 0 & 0
& 0 & \downarrow & \uparrow & 0 & \downarrow & 0 & \uparrow & \cdots \\
m_{2}: & \cdots & \uparrow & 0 & \downarrow & 0 & 0 & 0 & \uparrow &
\downarrow & 0 & 0 & 0 & \uparrow & 0 & \downarrow & 0 & \cdots%
\end{array}%
\end{equation*}%
where $(\uparrow ,0,\downarrow )$ represent $|m\rangle =(|1\rangle
,|0\rangle ,\left\vert -1\right\rangle )$. This dilute antiferromagnetic\
order is in analogy with the spin-$1$ valence bond solid (VBS) state in
terms of the $S^{z}$ quantum number, but here two quantum numbers are
associated with the Cartan generators $L^{12}$ and $L^{34}$. However,
\textit{such an intriguing feature is not enjoyed by the }$S=2$\textit{\ }%
VBS states of Affleck, Kennedy, Lieb, and Tasaki (AKLT) model. Actually, the
characterization scheme of the VBS states for $S\geq 2$ remains a
challenging open problem. However, the hidden order of all $SO(2l+1)$ matrix
product states can be fully identified in a systematic and compact form.

\subsection{Hidden order in the $SO(2l+1)$ matrix product state}

Now we are in a position to identify the hidden order in all the $SO(2l+1)$
matrix product state (\ref{eq:MPS}), which is inspired from the analysis of $%
SO(5)$ matrix product state. Since $SO(2l+1)$ is a rank-$l$ algebra, one can
always choose the mutually commuting Cartan generators as $%
\{L^{12},L^{34},\ldots ,L^{2l-1,2l}\}$. At each site, the quantum states are
classified by the eigenvalues of these Cartan generators as
\begin{equation}
L^{2\alpha -1,2\alpha }|m_{\alpha }\rangle =m_{\alpha }|m_{\alpha }\rangle
,\quad (m_{\alpha }=0,\pm 1).
\end{equation}%
Thus, the single-site states are associated with $l$ quantum numbers $%
\{m_{1},\cdots ,m_{l}\}$ and they are subjected to the constraint
\begin{equation}
m_{\alpha }m_{\beta }=0,\qquad (\alpha \not=\beta ).  \label{eq:constraint}
\end{equation}%
According to Eq. (\ref{Rotation}), all the Cartan generators annihilate the
\textquotedblleft extra dimension\textquotedblright\ vector $\left\vert
n^{2l+1}\right\rangle =\left\vert 0,0,\ldots ,0\right\rangle $. The other
basis states can be chosen as
\begin{equation}
\left\vert 0\ldots ,m_{\alpha }=\pm 1,\ldots 0\right\rangle =\frac{1}{\sqrt{2%
}}\left( \left\vert n^{2\alpha }\right\rangle \pm i\left\vert n^{2\alpha
-1}\right\rangle \right) .  \label{eq:CartanBasis}
\end{equation}

From the property of the Clifford algebra, the hidden antiferromagnetic
order of the ground state $|\Psi \rangle $ can now be identified. In any of
the $m_{\alpha }\,(\alpha =1\sim l)$ channel, it can be shown that $%
|m_{\alpha }\rangle $ is diluted antiferromagnetically ordered, the same as
for the $S=1$ VBS state. Namely, the states of $m_{\alpha }=1$ and $-1$ will
alternate in space if all the $m_{\alpha }=0$ states between them are
ignored.

This hidden antiferromagnetic order can also be characterized by nonlocal
string order parameters. Similar to the $l=1$ case, the string order
parameters can be defined as
\begin{equation}
\mathcal{O}^{ab}=\lim_{|j-i|\rightarrow \infty }\langle
L_{i}^{ab}\prod_{r=i}^{j-1}\exp (i\pi L_{r}^{ab})L_{j}^{ab}\rangle .
\label{eq:SOP}
\end{equation}%
Since the ground state is $SO(2l+1)$ rotationally invariant, the above
nonlocal order parameters should all be equal to each other. Thus, to
determine the value of these parameters, only $\mathcal{O}^{12}$ needs to be
evaluated. One can compute the value of these string order parameters by the
transfer-matrix techniques but there is an alternate intuitive approach. In
the $L^{12}$ channel, the role of the phase factor in Eq. (\ref{eq:SOP}) is
to correlate the finite spin-polarized states in the $m_{1}$ channel at the
two ends of the string. If nonzero $m_{1}$ takes the same value at the two
ends, then the phase factor is equal to $1$. On the other hand, if nonzero $%
m_{1}$ takes two different values at the two ends, then the phase factor is
equal to $-1$. Thus, the value of $\mathcal{O}^{12}$ is determined purely by
the probability of $m_{1}=\pm 1$ appearing at the two ends of the string. It
is straightforward to show that the probability of the states $m_{1}=\pm 1$
appearing at one lattice site is $2/(2l+1)$ and thus $\mathcal{O}%
^{12}=4/(2l+1)^{2}$.

\begin{figure}[tbp]
\includegraphics[scale=0.5]{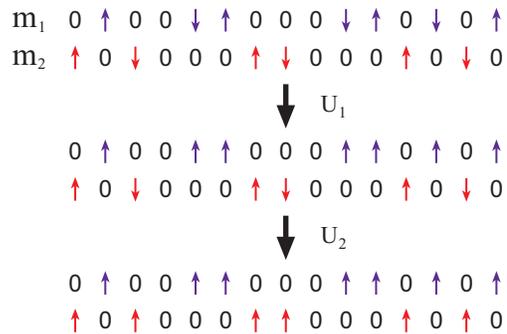}
\caption{(Color online) Changes of a typical configuration of the $SO(5)$
ground state under the unitary transformation defined by Eq. (\protect\ref%
{eq:transf}). $U_{1}$ and $U_{2}$ transform successively all $m_{1}$ and $%
m_{2}$ states to two diluted ferromagnetic configurations, respectively.}
\label{fig:transf}
\end{figure}

In the $SO(2l+1)$ Lie algebra, $(L^{2\alpha -1,2\alpha }$, $L^{2\alpha
-1,2l+1}$, and $L^{2\alpha ,2l+1})$ span an $SO(3)$ sub-algebra in which $%
\exp (i\pi L^{2\alpha ,2l+1})$ plays the role of flipping the quantum number
$m_{\alpha }$. This exponential operator can flip the quantum numbers of $%
m_{\alpha }$ without disturbing the quantum states in all other channels.
This indicates that if we take the following nonlocal unitary transformation
in the $m_{\alpha }$ channel:
\begin{equation}
U_{\alpha }=\prod_{j<i}\exp \left( i\pi L_{j}^{2\alpha -1,2\alpha
}L_{i}^{2\alpha ,2l+1}\right) ,
\end{equation}%
then all the configurations in this channel will be ferromagnetically
ordered. Furthermore, by performing this nonlocal transformation
successively in all the channels,
\begin{equation}
U=\prod_{\alpha =1}^{l}U_{\alpha },  \label{eq:transf}
\end{equation}%
then all the configurations of the ground state will become
ferromagnetically ordered. As an example, Fig. \ref{fig:transf} shows how a
typical configuration of the $SO(5)$ matrix product state is successively
changed under this nonlocal unitary transformation.

By applying the unitary transformation (\ref{eq:transf}) to the \textit{%
Cartan} generators, it can be shown that
\begin{equation}
UL_{i}^{ab}U^{-1}=L_{i}^{ab}\exp (i\pi \sum_{j=1}^{i-1}L_{j}^{ab}).
\end{equation}%
Substituting this formula to Eq. (\ref{eq:SOP}), we find that
\begin{equation}
\mathcal{O}^{ab}=\lim_{\left\vert j-i\right\vert \rightarrow \infty
}\left\langle L_{i}^{ab}L_{j}^{ab}\right\rangle _{U}.  \label{eq:CF}
\end{equation}%
Thus, the nonlocal string order parameters $\mathcal{O}^{ab}$ for Cartan
generators become the ordinary two-point correlation functions of local
operators after the unitary transformation.

Under the above transformation, the $SO(2l+1)$ symmetry of the original
Hamiltonian is reduced and determined by the symmetry of the unitary
transformation operators. In the $m_{\alpha }$ channel, it can be shown that
the unitary operator $U_{\alpha }$ possesses only a $Z_{2}\times Z_{2}$
symmetry. Therefore, the Hamiltonian after the transformation has a $%
(Z_{2}\times Z_{2})^{l}$ symmetry. This is the hidden topological symmetry
of the Hamiltonian, associated with the hidden order of the original matrix
product state $|\Psi \rangle $. When it is applied to an open chain system,
the hidden $(Z_{2}\times Z_{2})^{l}$ topological symmetry of the Hamiltonian
will be further broken, yielding $2^{l}$-free edge states at each end of the
chain. Therefore, the open chain has totally $4^{l}$-degenerate ground
states, which can be distinguished by their edge states.

\section{Even-$n$ case}

Let us assume $n=2l\ (l\geq 2)$. Using the $(2l+1)$ gamma matrices, the
spinor representation of the $SO(2l)$ algebra can be constructed by leaving
out $\Gamma ^{2l+1}$. However, we note that the resulting $2^{l}$%
-dimensional spinor representation generated by $\Gamma ^{a}$ $(a=1\sim 2l)$
is \textit{reducible}, in contrast to the $SO(2l+1)$ algebra. \cite%
{Georgi-1999} Since $\Gamma ^{2l+1}$ commutes with all the $SO(2l)$
generators $\Gamma ^{ab}$, one can construct the following projection
operators onto two different invariant subspaces:%
\begin{equation}
P_{\pm }=\frac{1}{2}(1\pm \Gamma ^{2n+1}).
\end{equation}

For each lattice site $j$, we introduce the local $g$ matrix as
\begin{equation}
g_{j}=\sum_{a=1}^{2l}\Gamma ^{a}\left\vert n^{a}\right\rangle _{j},
\end{equation}%
then the exact matrix product ground states of the Hamiltonian (\ref%
{eq:soham}) for $n=2l$ are given by
\begin{eqnarray}
|\Psi _{\pm }\rangle &=&\mathrm{Tr}\left( P_{\pm }g_{1}g_{2}\ldots
g_{N}\right)  \notag \\
&=&\sum_{a_{1}\ldots a_{N}}\mathrm{Tr}(P_{\pm }\Gamma ^{a_{1}}\ldots \Gamma
^{a_{N}})\left\vert n^{a_{1}}\ldots n^{a_{N}}\right\rangle .
\label{eq:SO(2l)MPS}
\end{eqnarray}%
Due to the equation $P_{\pm }\Gamma ^{a}=\Gamma ^{a}P_{\mp }$ $(a=1\sim 2l)$%
, we can observe that the states $|\Psi _{\pm }\rangle $ are dimerized
states and are connected to each other by translating one lattice site. Thus
these two states break translational symmetry while they preserve the $%
SO(2l) $ rotational symmetry. For an open chain, the matrix product ground
states are $2^{2l-1}$-fold degenerate when combining dimerization and edge
states.

The static correlation functions can be computed by the transfer-matrix
method as well. We find that the $SO(4)$ matrix product states have only
nearest-neighbor correlations $\left\langle
L_{i}^{ab}L_{i+1}^{ab}\right\rangle =-1/4$ and the correlation length is
zero. For $l\geq 3$, the two-point correlation function $\left\langle
L_{i}^{ab}L_{j}^{ab}\right\rangle $ has an exponential tail at a\ large
distance, as in Eq. (\ref{eq:decay}), and the correlation length is $\xi
=1/\ln (\frac{l}{l-2})$.

Although these two-point correlation functions of the $SO(2n)$ matrix
product states are short range, there is a hidden antiferromagnetic order,
similar to the $SO(2l+1)$ matrix product states. Because $SO(2l)$ is a rank-$%
l$ algebra, the Cartan generators can be chosen as $\{L^{12},L^{34},\ldots
,L^{2l-1,2l}\}$. Thus, the states can be characterized by the $SO(2l)$
weight using Eq. (\ref{eq:CartanBasis}). In the $SO(2l)$ case, the only
difference is the absence of the extra dimension\ vector $\left\vert
0,0,\ldots ,0\right\rangle $ annihilated by all Cartan generators. To
measure this hidden order, one can use the string order parameter in Eq. (%
\ref{eq:SOP}). A straightforward calculation shows the value of these string
order parameters that are given by%
\begin{equation}
\mathcal{O}^{ab}=\lim_{|j-i|\rightarrow \infty }\langle
L_{i}^{ab}\prod_{r=i}^{j-1}\exp (i\pi L_{r}^{ab})L_{j}^{ab}\rangle =\frac{1}{%
l^{2}}.  \label{eq:SO(2l)SOP}
\end{equation}

Here we note that translational symmetry breaking distinguishes the $SO(2l)$
case from the $SO(2l+1)$ case, while the latter belongs to the Haldane spin
liquid class. This is an interesting even-odd effect. Furthermore, one may
expect that their low-lying excitations are also very different. The
low-energy excitations in the $SO(2l+1)$ Haldane liquid are magnons, while
the $SO(2l)$ systems are soliton-like excitations connecting the two
dimerized states. Although the exact results of the low-lying excitations do
not exist, more evidence comes from the $SO(4)$ case.

\subsection{SO(4) matrix product state: A staggered spin-orbital crystal}

The $SO(4)$ case is somewhat special because it can be factorized as $%
SO(4)\simeq SU(2)\times SU(2)$ with $\mathcal{%
%TCIMACRO{\U{2102} }%
%BeginExpansion
\mathbb{C}
%EndExpansion
}^{2}\otimes \mathcal{%
%TCIMACRO{\U{2102} }%
%BeginExpansion
\mathbb{C}
%EndExpansion
}^{2}$ being its vector representation. Namely, one can consider a
spin-orbital $S=T=1/2$ coupled chain or equivalently a spin-$1/2$ two-leg
spin ladder to implement the $SO(4)$ vectors and generators in Eq. (\ref%
{Rotation}). We find that it is convenient to introduce the four vector
states as%
\begin{eqnarray}
\left\vert n^{1,2}\right\rangle &=&\frac{e^{\pm i\pi /4}}{\sqrt{2}}%
(\left\vert \uparrow ,\uparrow \right\rangle \mp \left\vert \downarrow
,\downarrow \right\rangle ),  \notag \\
\left\vert n^{3,4}\right\rangle &=&\frac{e^{\mp i\pi /4}}{\sqrt{2}}%
(\left\vert \downarrow ,\uparrow \right\rangle \mp \left\vert \uparrow
,\downarrow \right\rangle ),
\end{eqnarray}%
where the first index in $\left\vert \sigma ,\tau \right\rangle $ denotes
the spin direction while the second one is the orbital direction. Moreover,
the $SO(4)$ generators are defined by%
\begin{eqnarray}
L^{12} &=&-T^{z}-S^{z},\text{ }L^{13}=T^{x}-S^{x},\text{ }%
L^{14}=-T^{y}-S^{y},  \notag \\
L^{23} &=&T^{y}-S^{y},\text{ \ }L^{24}=T^{x}+S^{x},\text{ \ }%
L^{34}=T^{z}-S^{z},
\end{eqnarray}%
where $S^{\mu }$ and $T^{\mu }$ $(\mu =x$, $y$, or $z)$ denote the spin and
orbital degrees of freedom, respectively. Alternately, $S^{\mu }$ and $%
T^{\mu }$ can be viewed as the spin operators in the upper and lower leg of
a two-chain ladder.

A convenient choice of $\Gamma $ matrices for $SO(4)$ spinor representation
is given by
\begin{eqnarray}
\Gamma ^{1} &=&\sigma ^{2}\otimes \sigma ^{3},\text{ }\Gamma ^{2}=-\sigma
^{1}\otimes \sigma ^{3},  \notag \\
\Gamma ^{3} &=&\sigma ^{0}\otimes \sigma ^{2},\text{ }\Gamma ^{4}=-\sigma
^{0}\otimes \sigma ^{1}.
\end{eqnarray}
The invariant subspace projector is $P_{\pm }=(1\pm \Gamma ^{5})/2$ and $%
\Gamma ^{5}=\sigma ^{3}\otimes \sigma ^{3}$. A little calculation shows that
the local $g$ matrix is given by
\begin{equation}
g_{j}=%
\begin{pmatrix}
0 & -\left| \downarrow ,\uparrow \right\rangle _{j} & i\left| \downarrow
,\downarrow \right\rangle _{j} & 0 \\
-\left| \uparrow ,\downarrow \right\rangle _{j} & 0 & 0 & -i\left|
\downarrow ,\downarrow \right\rangle _{j} \\
i\left| \uparrow ,\uparrow \right\rangle _{j} & 0 & 0 & -\left| \downarrow
,\uparrow \right\rangle _{j} \\
0 & -i\left| \uparrow ,\uparrow \right\rangle _{j} & -\left| \uparrow
,\downarrow \right\rangle _{j} & 0%
\end{pmatrix}%
,
\end{equation}%
up to an unimportant normalization factor.

\begin{figure}[tbp]
\includegraphics[scale=0.9]{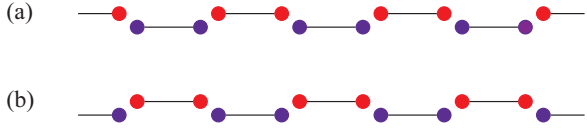}
\caption{(Color online) The schematic of the twofold-degenerate staggered
spin-orbital crystal states (a) $|\Psi _{+}\rangle $ and (b) $|\Psi
_{-}\rangle $.}
\label{fig:spinorbital}
\end{figure}

In this case, the twofold-degenerate ground states have an intuitive
meaning, which becomes clear when the local Hilbert space is represented by
a Schwinger-boson Fock space as $\left\vert \sigma ,\tau \right\rangle
=a_{\sigma }^{\dagger }b_{\tau }^{\dagger }\left\vert \mathrm{vac}%
\right\rangle $. Here $a_{\sigma }^{\dagger }$ and $b_{\tau }^{\dagger }$
create a state with spin and orbital directions $\sigma $ and $\tau $,
respectively. Using these Schwinger bosons, we find that the state $%
\left\vert \Psi _{+}\right\rangle $ in Eq. (\ref{eq:SO(2l)MPS}) can be
written as%
\begin{eqnarray}
\left\vert \Psi _{+}\right\rangle &=&\prod_{i=1}^{N/2}(b_{2i-1,\uparrow
}^{\dagger }b_{2i,\downarrow }^{\dagger }-b_{2i-1,\downarrow }^{\dagger
}b_{2i,\uparrow }^{\dagger })  \notag \\
&&\times (a_{2i,\uparrow }^{\dagger }a_{2i+1,\downarrow }^{\dagger
}-a_{2i,\downarrow }^{\dagger }a_{2i+1,\uparrow }^{\dagger })\left\vert
\mathrm{vac}\right\rangle ,
\end{eqnarray}%
and the interchange of $a^{\dagger }$ and $b^{\dagger }$ yields $\left\vert
\Psi _{-}\right\rangle $. These staggered spin-orbital crystal states are
first found by Kolezhuk and Mikeska. \cite{Kolezhuk-1998} The picture of
these states is displayed in Figs. \ref{fig:spinorbital}(a) and \ref%
{fig:spinorbital}(b). Obviously, the two-point correlation functions are
nonvanishing only between nearest-neighbor sites. For this spin-orbital $%
SO(4)$ system, the string order parameters in Eq. (\ref{eq:SO(2l)SOP}) can
be written as%
\begin{equation}
\lim_{|j-i|\rightarrow \infty }\langle \left( S_{i}^{\mu }\pm T_{i}^{\mu
}\right) \prod_{r=i}^{j-1}e^{i\pi (S_{r}^{\mu }\pm T_{r}^{\mu })}(S_{j}^{\mu
}\pm T_{j}^{\mu })\rangle =\frac{1}{4},
\end{equation}%
where $\mu =x$, $y$, or $z$. In the studies of two-leg spin ladders, these
types of string order parameters were introduced to divide the topologically
distinct gapped spin liquid states. \cite{EHKim-2000}

The fact that such two dimerized states are exact ground states of the
projector Hamiltonian (\ref{eq:soham}) can be easily visualized when writing
the projectors of the three $SO(4)$ channels in Eq. (\ref{eq:so(n)projector}%
) as
\begin{eqnarray}
\mathcal{P}_{\underline{1}}(i,j) &=&P_{S=0}(i,j)P_{T=0}(i,j),  \notag \\
\mathcal{P}_{\underline{6}}(i,j)
&=&P_{S=0}(i,j)P_{T=1}(i,j)+P_{S=1}(i,j)P_{T=0}(i,j),  \notag \\
\mathcal{P}_{\underline{9}}(i,j) &=&P_{S=1}(i,j)P_{T=1}(i,j),
\end{eqnarray}%
where $P_{S=0}(i,j)=\frac{1}{4}-\mathbf{S}_{i}\cdot \mathbf{S}_{j}$ and $%
P_{S=1}(i,j)=\frac{3}{4}+\mathbf{S}_{i}\cdot \mathbf{S}_{j}$ are bond total
spin projectors. Once the spin and orbital singlets are formed between
nearest-neighbor sites in a staggered pattern, the $SO(4)$-symmetric
projector Hamiltonian,%
\begin{eqnarray}
H_{SO(4)} &=&\sum_{i}\mathcal{P}_{\underline{9}}(i,i+1)  \notag \\
&=&\sum_{i}(\mathbf{S}_{i}\cdot \mathbf{S}_{i+1}+\frac{3}{4})(\mathbf{T}%
_{i}\cdot \mathbf{T}_{i+1}+\frac{3}{4}),
\end{eqnarray}%
always annihilate such a spin-orbital crystal state.

\section{$SO(n)$ bilinear-biquadratic model}

As already mentioned, $H_{SO(n)}$ is a bilinear-biquadratic Hamiltonian in
terms of the $SO(n)$ generators. More generally, we can also introduce a
one-parameter family of the $SO(n)$ symmetric bilinear-biquadratic model,
\begin{equation}
H_{bb}=\sum_{i}\left[ \cos \theta \sum_{a<b}L_{i}^{ab}L_{i+1}^{ab}+\sin
\theta \left( \sum_{a<b}L_{i}^{ab}L_{i+1}^{ab}\right) ^{2}\right] ,
\label{eq:BB}
\end{equation}%
which is an extension of the familiar spin-$1$ bilinear-biquadratic model.
The absence of the higher-order terms follows from the fact that such terms
can be expressed via the lower-order terms by means of Eq. (\ref%
{eq:so(n)projector}). To sketch the properties of this bilinear-biquadratic
model, we need to identify several special integrable points. Let us
introduce a slave boson representation,%
\begin{equation}
\left\vert n^{a}\right\rangle =d_{a}^{\dagger }\left\vert \mathrm{vac}%
\right\rangle ,
\end{equation}%
which yields a constraint $\sum_{a=1}^{n}d_{a}^{\dagger }d_{a}=1$. Using the
slave bosons, the $SO(n)$ generators can be written as $L^{ab}=i(d_{a}^{%
\dagger }d_{b}-d_{b}^{\dagger }d_{a})$, and the $SO(n)$-singlet bond
projector in Eq. (\ref{eq:so(n)projector}) is given by

\begin{equation}
\mathcal{P}_{\underline{1}}(i,j)=\frac{1}{n}\sum_{ab}d_{ia}^{\dagger
}d_{ja}^{\dagger }d_{jb}d_{ib}.
\end{equation}%
Additionally, the $SU(n)$-invariant permutation operator is expressed as%
\begin{equation}
Q(i,j)=\sum_{ab}d_{ia}^{\dagger }d_{ib}d_{jb}^{\dagger }d_{ja}.
\end{equation}%
Using the permutation operator and the singlet projector, we can express the
bilinear-biquadratic Hamiltonian (\ref{eq:BB}) as%
\begin{eqnarray}
H_{bb} &=&\sum_{i}\{\cos \theta Q(i,i+1)  \notag \\
&&+n[(n-2)\sin \theta -\cos \theta ]\mathcal{P}_{\underline{1}}(i,i+1)\},
\end{eqnarray}%
up to a constant. Several special points can be identified as follows:

(1) $\theta =\tan ^{-1}\frac{1}{n-2}$ and $\tan ^{-1}\frac{1}{n-2}-\pi $;
the Hamiltonian (\ref{eq:BB}) reduces to a sum of nearest-neighbor
permutation operators, thus, it has an enhanced $SU(n)$ symmetry. In this
case, the transformation on each lattice site is in the $SU(n)$ fundamental
representation. For $\theta =\tan ^{-1}\frac{1}{n-2}$, this is the
Uimin-Lai-Sutherland (ULS)\ model, \cite{Sutherland-1975} which can be
solved by Bethe-ansatz method. It is known that there are gapless
excitations above the ground states and the effective low-energy field
theory is described by an $SU(n)_{1}$ Wess-Zumino-Witten model. \cite%
{Affleck-1986}

(2) $\theta =\pm \frac{\pi }{2}$; the Hamiltonian (\ref{eq:BB}) reduces to a
sum of nearest-neighbor singlet projectors and also has an $SU(n)$ symmetry.
However, the transformations are in the $SU(n)$ fundamental and its
conjugate representations on the even and odd numbers of lattice sites,
respectively. For $\theta =-\frac{\pi }{2}$, a mapping to the $n^{2}$-state
quantum Potts model allows the model to be solved exactly, \cite%
{Parkinson-1987,Barber-1989,Klumper-1989} and the ground states are
dimerized states with a finite-energy gap.

\begin{figure}[tbp]
\includegraphics[scale=1]{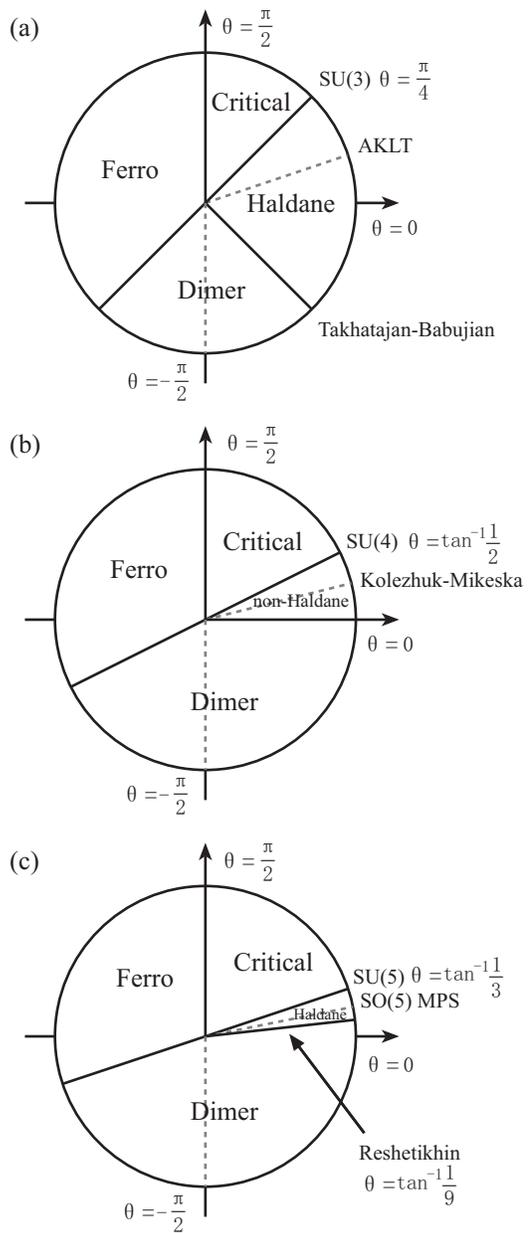}
\caption{The ground-state phase diagrams of the (a) $SO(3)$, (b) $SO(4)$,
and (c) $SO(5)$ symmetric bilinear-biquadratic spin chains.}
\label{bbFig}
\end{figure}

(3) $\theta =\tan ^{-1}(n-4)/(n-2)^{2}$; the Hamiltonian (\ref{eq:BB}) was
exactly solved by Reshetikhin \cite{Reshetikhin-1983}\ via quantum inverse
scattering method, which also exhibited gapless excitations. For $n=3$, this
point corresponds to the spin-$1$ Takhatajan-Babujian model, \cite%
{Takhatajan-1982} which is the quantum critical point between Haldane gap
phase and dimerized phase. For $n=4$, the Reshetikhin point yields the $%
SO(4) $ Heisenberg model, which is equivalent to two \textit{decoupled} spin-%
$1/2$ Heisenberg antiferromagnetic spin chains.

(4) $\theta =\tan ^{-1}\frac{1}{n}$; the ground states of the model
Hamiltonian (\ref{eq:BB}) are just the matrix product states considered in
Secs. III and IV. For an odd $n$, the ground state is a unique Haldane
liquid state. For an even $n$, the ground states are twofold-degenerate
dimerized states and are referred to non-Haldane liquid states.

Therefore, these rigorous results suggest that an energy gap develops for
the model (\ref{eq:BB}) in the finite parameter region%
\begin{equation}
\tan ^{-1}\frac{n-4}{(n-2)^{2}}<\theta <\tan ^{-1}\frac{1}{n-2},
\label{eq:GapRegion}
\end{equation}%
which always includes our matrix product ground-state point $\theta =\tan
^{-1}\frac{1}{n}$. The gap formation in this region is quite subtle. In the
point of view of conformal field theory, the $SU(n)$ ULS and the $SO(n)$
Reshetikhin points are both conformal invariant and are characterized by two
effective-field theories with different central charges. If so, there will
be no renormalization flow from the $SO(n)$ Reshetikhin point to the $SU(n)$%
-symmetric point according to Zamolodchikov's $c$ theorem, \cite%
{Zamolodchikov-1986} and an energy gap must be generated between these two
conformal invariant points. It was known that the conformal field theory for
the ULS point is an $SU(n)_{1}$ Wess-Zumino-Witten model with central charge
$c=n-1$. The conformal field theory description for the $SO(n)$ Reshetikhin
point is an $SO(n)_{1}$ Wess-Zumino-Witten model with central charge $c=n/2$%
. In particular, the $SO(3)$ Takhatajan-Babujian model is known to have a
central charge $c=3/2$ and the $SO(4)$ Heisenberg model (two decoupled spin-$%
1/2$ Heisenberg antiferromagnetic spin chains) has a central charge $c=2$.

Toward the odd $n=2l+1$ case, Itoi and Kato \cite{Itoi-1997}\ found that a
marginally relevant perturbation around the ULS point develops a Haldane gap
for $\theta <\tan ^{-1}\frac{1}{n-2}$, while the region $\theta >\tan ^{-1}%
\frac{1}{n-2}$ near the ULS point is massless. However, the even $n$ case
was extensively studied for $n=4$, which corresponds to an $SO(4)$
spin-orbital coupled system. \cite%
{Nersesyan-1997,Pati-1998,Azaria-1999,Itoi-2000,GMZhang-2003} These results
reveal that there is a dimerized non-Haldane liquid phase with an energy gap
between the $SU(4)$-symmetric point and the $SO(4)$ Heisenberg point. In the
non-Haldane liquid state, magnon excitations are incoherent and the
low-energy excitations are a pair of solitons connecting two spontaneously
dimerized ground states. It is thus expected that these $SO(n)$ symmetric
models also show such an interesting even-odd effect not only in the ground
states but also in the low-energy excitations. For $n=2l+1$, the system is
in a Haldane gap liquid phase with magnon excitations. For $n=2l$, the
elementary excitations are solitons connecting the degenerate ground states.

Following the exact results, the main phase diagrams of $SO(n)$
bilinear-biquadratic model for $n=3,4,5$ are displayed in Fig. (\ref{bbFig}%
). However, the $SO(n)$ antiferromagnetic Heisenberg model deserves more
attention, corresponding to the bilinear-biquadratic model (\ref{eq:BB})
with a pure bilinear interaction for $\theta =0$. When $n=3$, it is just the
quantum spin-$1$ antiferromagnetic Heisenberg model, which is in the Haldane
gap region. When $n=4$, the $SO(4)$ Heisenberg model is equivalent to two
decoupled spin-$1/2$ antiferromagnetic chains, which have unique disordered
ground states with power-law decay spin correlations. However, when $n=5$,
we find that the $SO(5)$ \textit{antiferromagnetic Heisenberg model is not
included in the Haldane gap region}. Therefore, it is interesting to ask
what are the ground states of the $SO(n)$ antiferromagnetic models for $%
n\geq 5$. Based on a generalized Lieb-Schultz-Mattis theorem, Li \cite%
{YQLi-2001}\ studied $SO(n)$ antiferromagnetic models for $n=4,5,6$. He
found that the $SO(4)$ Heisenberg model is gapless, while $SO(5)$ and $SO(6)$
Heisenberg models are suspected to have a gap. Together with our results, we
predict that the $SO(n)$ Heisenberg model for $n\geq 5$ belongs to the
dimerized phase with a finite-energy gap.

\section{Conclusion}

In conclusion, we have introduced a class of $SO(n)$ symmetric spin chain
Hamiltonians with nearest-neighbor interactions, whose exact ground states
are two different $SO(n)$ symmetric matrix product states depending on the
parity of $n$.

For an odd $n=2l+1$, a periodic chain has a unique ground state, which
preserves an $SO(2l+1)$ rotational and translational symmetries. The $%
SO(2l+1)$ symmetric spin chains with different $l$ are directly related to
quantum integer-spin chains belonging to the Haldane gap phase with a hidden
antiferromagnetic order characterized by nonlocal string order parameters.
The hidden $(Z_{2}\times Z_{2})^{l}$ symmetry responsible for the hidden
order has been found by applying a unitary transformation to the model
Hamiltonian. The Haldane gap and $4^{l}$ degenerate ground states in an open
chain are natural consequences of this hidden symmetry breaking.

For an even $n=2l$, a periodic chain has a twofold-degenerate dimerized
ground state, which preserves $SO(2l)$ symmetry but breaks translational
symmetry. These $SO(2l)$ matrix product states with different $l$ are
non-Haldane liquid states, which have soliton excitations connecting the two
degenerate ground states. However, these $SO(2l)$ matrix product states also
contain a hidden antiferromagnetic order characterized by nonlocal string
order parameters.

Finally, a generalized $SO(n)$ symmetric bilinear-biquadratic model family
has been discussed and the ground-state phase diagrams are sketched based on
some known exact results. One of the important conclusions is that the
ground state of the $SO(n)$ symmetric Heisenberg antiferromagnetic spin
model for $n\geq 5$ is predicted to be in a twofold-degenerate dimerized
state. Further investigations on this are certainly required.

We acknowledge the support of NSF of China and the National Program for
Basic Research of MOST-China.

\end{document}